\documentclass[12pt]{article}
\usepackage{epsfig}
\usepackage{amssymb}
\usepackage{amsmath}
\usepackage{amsfonts}

\newcommand{\C}{\mathbb{C}}

\newcommand{\be}{\begin{equation}}
\newcommand{\ee}{\end{equation}}
\newcommand{\bea}{\begin{eqnarray}}
\newcommand{\eea}{\end{eqnarray}}

\newcommand{\kt}{\rangle}
\newcommand{\br}{\langle}

\newcommand{\ed}{\end{document}}

\begin{document}

\title{Comment on ``On Existence of a Biorthonormal
Basis Composed of Eigenvectors of  Non-Hermitian\\ Operators
[quant-ph/0603075]''}

\author{\\
Ali Mostafazadeh
\\
\\
Department of Mathematics, Ko\c{c} University,\\
34450 Sariyer, Istanbul, Turkey\\ amostafazadeh@ku.edu.tr}
\date{ }
\maketitle

\begin{abstract}
We point out that T.~Tanaka's recent criticism [quant-ph/0603075]
of the results of J.~Math.\ Phys. {\bf 43}, 3944 (2002)
[math-ph/0203005] is based on an assumption which was never made
in the latter paper, namely that the diagonalizability of an
operator implies that it is normal. Therefore, Tanaka's objections
regarding this paper are not valid.



\end{abstract}

The definition of a diagonalizable operator acting in an
infinite-dimensional vector space is not usually given in
textbooks on linear algebra or operator theory. This was the
reason I hesitated to use this term in \cite{p1,p2} and indeed it
was upon the insistence of the referee of \cite{p3}, who had found
the repeated statement of the assumption of the ``existence of a
complete biorthonormal basis of eigenvectors'' too lengthly, that
I decided to use the term ``diagonalizable'' in \cite{p3}.
According to \cite{p3,p6}, a linear operator $H$ acting in a
separable Hilbert space and having a discrete spectrum is called
\emph{diagonalizable} if there are eigenvectors $\psi_n$ of $H$
and $\phi_n$ of $H^\dagger$ that form a complete biorthonormal
basis (or system) $\{\psi_n,\phi_n\}$, i.e., they satisfy
    \be
    \br\psi_n|\phi_m\kt=\delta_{mn},~~~~~~~
    \sum_n|\psi_n\kt\br\phi_n|=\sum_n|\phi_n\kt\br\psi_n|=1.
    \label{biorth}
    \ee
Nowhere in this definition is it assumed that the operator is
normal. A normal operator, in finite-dimensions with no extra
conditions and in infinite-dimensions with appropriate extra
conditions, admit a diagonal matrix representation in some
orthonormal basis. This is usually called ``diagonalizability by a
unitary transformation.'' The operators considered in \cite{p3}
are not assumed to be normal operators and the related remarks of
Tanaka \cite{tanaka} do not hold. For example, according to the
above definition of a diagonalizable operator that is used in
\cite{p3}, the operator $A:\C^2\to\C^2$ that is represented in the
standard basis of $\C^2$ by the matrix
    \[\left(\begin{array}{cc}
    1 & 2 \\
    0 & 3\end{array}\right)\]
is certainly diagonalizable, for the following vectors (also
represented in the standard representation of $\C^2$) form a
nontrivial biorthonormal basis $\{\psi_n,\phi_n\}$ satisfying all
the above-mentioned properties.
    \[\psi_1=\left(\begin{array}{c}
    1  \\
    0\end{array}\right),~~~~~
    \psi_2=\left(\begin{array}{c}
    1  \\
    1\end{array}\right),~~~~~
    \phi_1=\left(\begin{array}{c}
    1  \\
    -1\end{array}\right),~~~~~
    \phi_2=\left(\begin{array}{c}
    0  \\
    1\end{array}\right).\]
This example clearly contradicts Tanaka's following claim
\cite{tanaka}: ``\emph{the operators which satisfy the assertion
stated in Ref.~[14]'' (which is Ref.~\cite{p3} of the present
paper) ``with a non-trivial biorthonormal basis must fall into, if
exist, a quite peculiar class of linear operators. It should be
noted in particular that there is no such operators in
finite-dimensional spaces.}'' This claim seems to be the result of
the misunderstanding that diagonalizable operators are necessarily
normal, an assumption which has never been made in \cite{p3,p6}
and other papers of mine and as far as I know of others on the
subject.

In conclusion, I should like to emphasize that in a quantum theory
fulfilling the measurement axiom the diagonalizability (as defined
above and in \cite{p3}) of the operators that are to be identified
with the observables is an absolutely necessary condition
\cite{jpa-2004b}. The situation is just the opposite for normal
operators that fail to be self-adjoint. Such operators have
non-real spectra and cannot be employed as observables in a
unitary quantum system. A detailed discussion of diagonalizable
operators and biorthonormal systems emphasizing the mathematically
rigorous results as well as the physical relevance will be given
in \cite{review}.

\ed

{

}

\ed
\begin{thebibliography}{9}
\bibitem{p1} A.\ Mostafazadeh, J.\ Math.\ Phys.\ {\bf 43}, 205 (2002)
\bibitem{p2} A.~Mostafazadeh, J.\ Math.\ Phys.\ {\bf 43}, 2814 (2002)
\bibitem{p3} A.~Mostafazadeh, J.\ Math.\ Phys.\ {\bf 43}, 3944 (2002)
\bibitem{p6} A.~Mostafazadeh, J.\ Math.\ Phys.\ {\bf 43}, 6343 (2002),
Erratum-ibid. {\bf 44}, 943 (2003)
\bibitem{tanaka} T.~Tanaka, quant-ph/0603075
\bibitem{jpa-2004b} A.~Mostafazadeh and A.~Batal,
J.~Phys.~A: Math.\ Gen.\ {\bf 37}, 11645 (2004)
\bibitem{review} A.~Mostafazadeh, in preparation.
\end{thebibliography}

\begin{thebibliography}{99}
\bibitem{bcm} C.~M.~Bender, J.-H.~Chen, and K.~A.~Milton,
J.~Phys.~A {\bf 39}, 1657 (2006), hep-th/0511229
\bibitem{cjp-2004b} A.~Mostafazadeh, Czech J. Phys. {\bf 54},
1125 (2004), quant-ph/0407213
\bibitem{jpa-2005b} A.~Mostafazadeh, J.~Phys.~A {\bf 38}, 6557 (2005),
quant-ph/0411137
\bibitem{jones-jpa-2005} H.~F.~Jones, J.~Phys.~A {\bf 38}, 1741
(2005), quant-ph/0411171
\bibitem{critique} A.~Mostafazadeh,  quant-ph/0310164
\bibitem{comment} A.~Mostafazadeh,  quant-ph/0407070
\bibitem{jpa-2004b} A.~Mostafazadeh and A.~Batal,
J.~Phys.~A: Math.\ Gen.\ {\bf 37}, 11645 (2004), quant-ph/0408132
\bibitem{jmp-2005} A.~Mostafazadeh, J.~Math.~Phys.\ {\bf 46},
102108 (2005), quant-ph/0506094
\bibitem{banerjee} A.~Banerjee, Mod.~Phys.\ Lett.~A {\bf 20},
3013 (2005), quant-ph/0502163
\bibitem{p67} A.~Mostafazadeh,  quant-ph/0508195
\bibitem{bagchi-q-r} B.~Bagchi, C.~Quesne, and R.~Roychoudhury,
J.~Phys.~A {\bf 39}, L127 (2006), quant-ph/0511182
\bibitem{bjr} C.~M.~Bender, H.~F.~Jones, and R.~J.~Rivers, Phys.\
Lett.~B {\bf 625}, 333 (2005), hep-th/0508105
\bibitem{jones-2006} H.~Jones and J.~Mateo, quant-ph/0601188
\bibitem{dorey} P.~Dorey, C.~Dunning, and R.~Tateo,
J.~Phys.~A {\bf 34}, 5679 (2001), hep-th/0103051;\\
K.~C.~Shin, Commun.\ Math.\ Phys.\ {\bf 229}, 543 (2002),
math-ph/0201013
\end{thebibliography}
